\documentclass[10pt,aps,prb,twocolumn,showpacs]{revtex4-1}

\usepackage[english]{babel}

\usepackage{amsmath}    
\usepackage{mathtools}  
\usepackage{isomath}    %
\usepackage{amssymb}    
\usepackage{mathrsfs}   
\usepackage{textcomp}   
\usepackage{accents}    

\usepackage{graphicx}   
\usepackage{epstopdf}   
\usepackage{subfigure}  

\usepackage{makecell}
\usepackage{array}      
\usepackage{tabularx}   
\usepackage{multirow}   
\usepackage{booktabs}   
\usepackage{dcolumn}    

\usepackage{gensymb}    

\usepackage{setspace}

\usepackage[usenames,dvipsnames]{color} 
\usepackage{url}        
\usepackage{hyperref}   

\hypersetup{
  colorlinks,
  citecolor=Violet,
  linkcolor=Red,
  urlcolor=Blue}

\newcommand{\angstrom}{$\,$\AA{}}

\newcommand{\MoSTwo}{MoS$_2$}

\definecolor{myred}{rgb}{1,0,1}
\definecolor{mygreen}{rgb}{0,1,0}
\definecolor{myblue}{rgb}{0,0,1}
\definecolor{mypink}{rgb}{0.8,0,0.8}

\graphicspath{{./}}   

\begin{document}
\setlength{\heavyrulewidth}{0.08em}
\setlength{\lightrulewidth}{0.05em}
\setlength{\cmidrulewidth}{0.03em}
\setlength{\belowrulesep}{0.65ex}
\setlength{\belowbottomsep}{0.00pt}
\setlength{\aboverulesep}{0.40ex}
\setlength{\abovetopsep}{0.00pt}
\setlength{\cmidrulesep}{\doublerulesep}
\setlength{\cmidrulekern}{0.50em}
\setlength{\defaultaddspace}{0.50em}
\setlength{\tabcolsep}{3pt}

\title{Quantum conductance of MoS\texorpdfstring{$_2$}{2} armchair nanoribbons}

\author{F.~Tabatabaei}
\author{I.~Abdolhosseini~Sarsari$^{*}$}
\author{N.~Rezaei}
\author{M.~Alaei}
\affiliation{Department of Physics, Isfahan University of 
Technology, Isfahan, 84156-83111, Iran\\
\\
}
\begin{abstract}

Molybdenum disulfide (\MoSTwo{}) is layered transition-metal dichalcogenide (TMDC), which in its monolayer form, has the direct bandgap of 1.8 eV.
We investigated the effect of width and strain
on quantum transport for \MoSTwo{} armchair nanoribbons.
That indicates \MoSTwo{} 
armchair nanoribbons are a good candidate for transistors even with strain. 
\end{abstract}
\maketitle


\section{Introduction}
\label{sec:introduction}

In recent decades, one-dimensional (1D) nanostructures have been extensively investigated 
because of their interesting properties and their mesoscopic physics. 
Among them, carbon nanostructures are well-studied, theoretically 
and experimentally\cite{zhou1994defects,rodriguez1995catalytic,panella2005hydrogen,
kaskhedikar2009lithium,smalley2003carbon,son2006energy,smalley2003carbon}. 
On the other hand, carbon-based nanostructures are not suitable for some applications because
of their small bandgap. 

Fortunately, a good alternative to carbon-based systems is a class of inorganic layered materials 
which exhibit different properties and also could be synthesized in the form of nanotubes and nanoribbons \cite{yoon2011good,radisavljevic2011single,Zhang2015,Radisavljevic2013,li2012enhanced}. 
These systems could use in a large variety of applications of one-dimensional systems. 
Molybdenum disulfide (\MoSTwo{}) is layered transition-metal di-calcogonide (TMDC) 
semiconductor that has attracted considerable interest because of its properties. 
In its bulk structure, an indirect bandgap of about 1.2 eV is observed, while in monolayer
form, the bandgap increases to 1.8 eV \cite{Kuc2011} with a transition to direct bandgap, which makes
it useful for some applications in electronics, optoelectronics, and photovoltaic devices{\cite{Radisavljevic2013}}. 

It has already been shown that monolayer of \MoSTwo{} is an ideal material for 
valleytronics \cite{Zeng2012}. This fact is due to the inversion-symmetry breaking together with
 spin-orbit coupling leads to coupled spin and valley degree of freedoms in a monolayer 
 of \MoSTwo{} and other group-VI dichalcogenides. 
It means that it is possible to control both spin and valley in these 2D materials. 
It has been shown that Mo and W dichalcogenides can exhibit two thermodynamically stable hexagonal (H) and
tetragonal (T) structural phases which provide opportunities for flexible, low power and transparent
electronic devices \cite{duerloo2014structural}. 
It has been shown that zigzag nanoribbons are metallic and magnetic and armchair
nanoribbons are semiconductor and nonmagnetic \cite{Qi2013,Pan2012}. 
The magnetic state of zigzag nanoribbons becomes more stable after passivation 
by hydrogen \cite{Qi2013}. Passivation by different nonmetal atoms has also been 
investigated for armchair nanoribbons. It turns out that the most stable structure is 
obtained when Mo and S atoms are passivated by oxygen and hydrogen, respectively. 

In this paper, we focus on armchair nanoribbons. We investigated the effect of width and strain
on quantum transport for \MoSTwo{} armchair nanoribbons.   
Due to flexibility and the large band gap of \MoSTwo{} nanostructures we investigated the effect of 
strain on the quantum transport of our structures\cite{duerloo2014structural}. That indicates \MoSTwo{} 
armchair nanoribbons are the good candidate for transistors even with strain. 

\section{Computational details}
\label{sec:computational_details}

The first-principle calculations are performed utilizing the projector augmented
wave (PAW) pseudopotentials in the framework of density functional theory (DFT),
as implemented in the Quantum-Espresso package\cite{giannozzi2009quantum}. 
The exchange-correlation effects are evaluated using the generalized gradient 
approximation as proposed by Perdew-Bruke-Ernzerhof (PBE-GGA). 
The Monkhorest-Pack scheme of kpoint sampling is used for integration over the
first Brillouin zone. 
For monolayer and nanoribbons structures, $7\times7\times1$ and $1\times7\times1$ 
k-point meshes are used, respectively. 
The energy and the wave-function cut-offs are set to 400 Ry and 35 Ry, respectively. 
We set all vacuum at least to 12 \angstrom{}. The wannierization is obtained by Marzari 
and Vanderbilt's method using the WANNIER90 code \cite{wannier90,Marzari1997,Marzari2012,Marzari2003,Mostofi2008}. 
In the case of entangled energy, we used Souza, Marzari, and Vanderbilt approach \cite{Souza2001}. 
\section{Results and disscution}
\label{sec:results}
\textbf{Electronic Properties}: 

Firstly, we investigated the electronic structure of \MoSTwo{} monolayer. 
The optimized lattice constant was 3.16 \angstrom{} and the band gap obtained 1.76 eV, in 
good agreement with the previous calculation\cite{Kuc2011}.
Then we investigated MLWF (Maximally Localized Wannier Function)
of the monolayer. We chose d orbital of Mo atom and p orbital of S atoms for 
initial projections in wannierization and we obtained MLWF for monolayer, in coincident 
with the previous results\cite{shi2013quasiparticle}. 
\begin{figure}[t!]
 \includegraphics[width=0.50\textwidth]{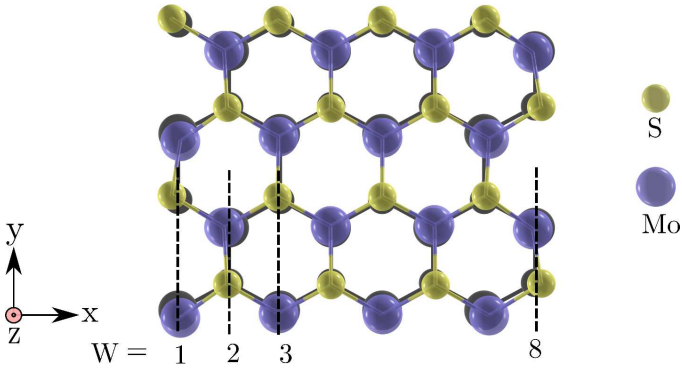}
 \caption{Optimized structure of a \MoSTwo{} 8-ANR, where the gray atoms in
background indicate the atomic positions before relaxation. 
The structure is periodic along the y direction.\label{bare-eps}} 
\end{figure}

After monolayer, we investigated \MoSTwo{} nanoribbons. Such as 
geraphene nanoribbons, \MoSTwo{} nanoribbons also can be described with 2 integer indices. 
So the chiral vector is  $\vec{C}_h= n \vec{a}+m \vec{b}$ that causes we have 2
types of nanoribbons: zigzag for $n\neq0 , m=0$ and armchair for $m=n$. 
From previous calculations, we know that zigzag nanoribbons are metallic 
and magnetic and armchair nanoribbons are semiconductor and 
nonmagnetic\cite{Qi2013}. Since our purpose is how to make a transistor, we 
chose armchair nanoribbons.

\begin{figure}[b]
 \includegraphics[scale=0.35]{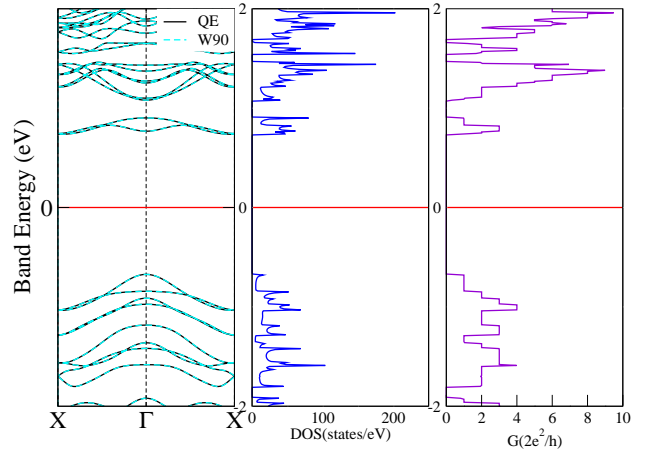}
 \caption{Band structure, density of state and conductance 
 of a the nanoribbons with 8 layer.\label{band8}}
 \end{figure}

\begin{figure}[t!]
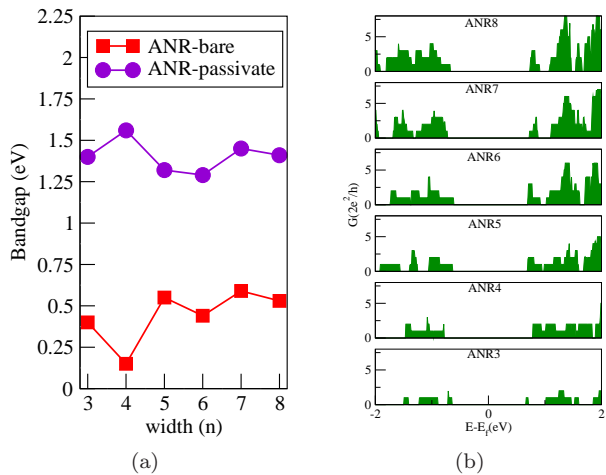

   \centering
   \subfigure[][]{
      \label{gap}
      \includegraphics*[width=0.43\linewidth]{gap3-8n.eps}}
   \hspace{0.02\textwidth}
   \subfigure[][]{
      \label{trans3-8}
      \includegraphics[width=0.405\linewidth]{trans3-8.eps}}
      \caption{ a) Energy band gap as a function of the ribbon width w with 
 3 $\preceq w \preceq 8$ for  A\MoSTwo{}NR-bare and  A\MoSTwo{}NR-H-O.
 b)The Landauer quantum conductance of an
A\MoSTwo{}NR-H-O in the y direction for w= 3- 8
 }
\end{figure}

After cutting monolayer and building armchair nanoribbons. 
we optimized lattice constant. Then the electronic spectrum of relaxed structures  
extracted from band structure calculation (see Fig.~\ref{band8} and Fig.~\ref{gap}). 
Others also reported the same results \cite{Qi2013}.

\begin{figure}[th]
 \includegraphics[width=0.50\textwidth]{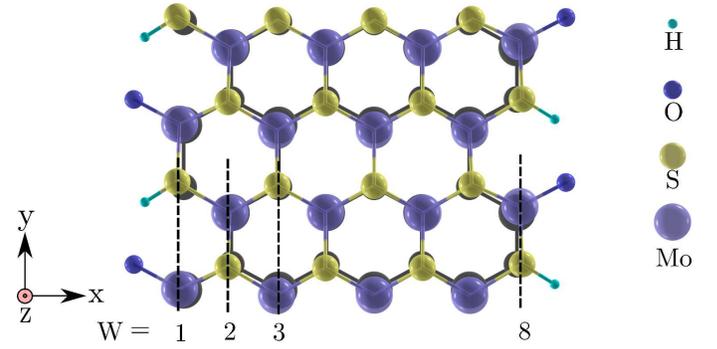}
 \caption{Optimized structure of a \MoSTwo{} 8-ANR-passivated, where the gray atoms in
background indicate the atomic positions before relaxation. 
The structure is periodic along the y direction.\label{pass}} 
\end{figure}

Then we modulated our structures with O and H atoms. Based on Zhang et. al paper
\cite{Zhang2015}, if each edge S atom is saturated by one H atom and each edge Mo atom is 
saturated by one O atom. 
As shown in (Fig.\ref{bare-eps}) and (Fig.\ref{pass}) in the case of ANR-bare the edge atoms after 
relaxation are more displaced than the case of ANR-passivated so passivation cause order in structure. 
We can see the bandgap of A\MoSTwo{}NR-H-Os 
converges to 1.4 eV that closer to the bandgap of monolayer compare to 
bare-nanoribbons (Fig.~\ref{pass}), and then similarity to bare-nanoribbons 
we obtained band structure of passivated nanoribbons.

Then binding energy per atom has been defined as:
 \begin{equation}
  E_{Binding~energy}=(E_{H-O}-E_{bare}-mE_H-nE_O)/(n+m)
 \end{equation}
 and computed for each width. 
 In table (\ref{table}) you can see bond length for edge atom and binding energies for every width. 
You can see our results for 3-A\MoSTwo{}NR to 8-A\MoSTwo{}NR (see Fig.~\ref{gap}). 
Our result is in good agreement with the previous result \cite{Zhang2015}. 
The bandgap is oscillating like the previous study but there is an interesting 
point in 4-ANR. 
\begin{table*}[ht]
\centering
\caption{Bond Distances (\angstrom{}) at the Edge
of Nanoribbons, Binding Energies E$_b$(eV), and Band Gaps E$_g$(eV) of Different AMoS$-2$NRs\label{table}}
 \begin{tabular}{|c| c |c| c| c| c |c|}
    \hline
 W&d$_{Mo-O}$\angstrom{} &d$_{S-H}$\angstrom{} &d$_{S-Mo}$\angstrom{} &E$_b$(eV)&E$_{g(bare)}$(eV)&E$_{g(passivated)}$(eV)\\
 \hline
 \hline
 3&1.70 &1.36&2.56 &-3.99& 0.40 &1.40   \\
 4& 1.70&1.35&2.37 &-4.11& 0.15 &1.56 \\
 5&1.71 &1.36&2.56 &-4.10& 0.55 &1.32 \\
 6&1.71 &1.36&2.37 &-4.11&  0.44&1.29 \\
 7& 1.71 &1.36&2.56&-4.10& 0.59 &1.45 \\
 8& 1.71&2.57&2.37 &-4.10& 0.53 &1.41 \\
   \hline
\end{tabular}
\end{table*}

For optimization lattice constant of nanoribbons we should obtain the relation between total energy and lattice constant. 
Because our structure is periodic in the y-direction we plotted total energy in terms of lattice constant in the y-direction. 
For all structure, we get one minimum in the plot but for 4-A\MoSTwo{}NR-bare we got two minima.
Of course, the global minima was right and was in good agreement with other works
\cite{Fan2014}. But there was another minimum that should be investigated (Fig.~\ref{2minimum}). 
As you saw in Fig.~\ref{gap}, the graph of an increasing/decreasing of bandgap after passivation remains 
as the same in bare one except for 4-A\MoSTwo{}NR.  
We investigate this little subject in the app.~\ref{app1}.
\vspace{2mm}

 \textbf{MLWF and quantum transport:} 

 After that, we should obtain MLWF 
 for each nanoribbon. It should be noted that choosing some appropriate 
 begin projections for wannierization lead to meaningful localized orbitals. 
After obtaining MLWF of each nanoribbon we can investigate the quantum transport 
of our systems. You can see our results in Fig.~\ref{trans3-8}, as you see with 
increasing width of the A\MoSTwo{}NR-H-Os number of passing channel is increased 
and the value of G (the quantum transport for each energy) is increased too. 
 
It should be noted that only electrons at the $E_f$ (Fermi energy) playing 
a role in the quantum conductance. At finite temperature, every sub-band 
near the $E_f$ can contribute to the quantum
conductance. The current, for a finite bias voltage (V), is given by 
\cite{Bagwell1989}

\begin{equation}\label{1}
\begin{aligned}
 I{}& =\int_0^\infty \frac{d\epsilon}{e}[f(\epsilon+\mu)-f(\epsilon)]G(\epsilon) \\
& \simeq \int_0^\infty d \epsilon[\mu \frac{\partial f}{\partial \epsilon} ] G(\epsilon) \\
 & =V \int_0^\infty d \epsilon \frac{\partial f}{\partial \epsilon} G(\epsilon)
\end{aligned}
   \end{equation}
   
where f is the Fermi-Dirac distribution function and $\mu$ is the chemical potential. 
At a finite temperature $\mu \frac{\partial f}{\partial \epsilon}$ has a Gaussian peak 
that at $T = 0$ K becomes a Dirac delta function. 
We can estimate that at room temperature the current can be approximated by 
integrating over $\epsilon \in [-0.3, 0.3]$ eV \cite{PhysRevB.81.125409} and 
compute conductance room temperature.

\begin{figure}[b!]
 \includegraphics[scale=0.8]{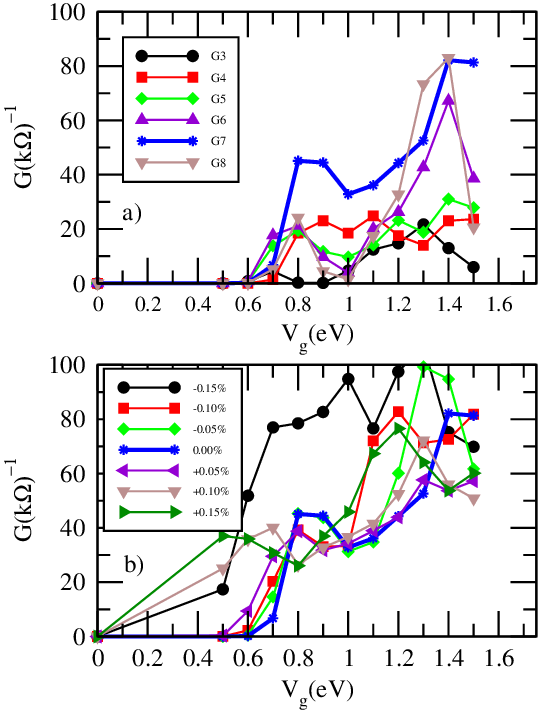}
 \caption{a) Conductance of A\MoSTwo{}NR-H-Os for various gate voltage 
 b) Conductance of 7-A\MoSTwo{}NR-H-Os for various strain and gate voltage
 \label{G-W-S}}
\end{figure}

Because of bandgap presence in A\MoSTwo{}NR-H-O, we can use them in transistors. 
As a result, we should check how the quantum transport change if the system influence
by various gate voltage (see Fig.~\ref{G-W-S}~a) As you see all of the nanoribbons are good 
for transistors because of their high ratio of On/Off in different gate voltages. 
But one should choose a proper gate voltage for transistors. 
For example, if you want to use 8-A\MoSTwo{}NR-H-O for the transistor using  
0.8 V and 1.1 V to 1.5 V value for gate voltage is recommended.

\begin{figure}[b]
 \includegraphics[scale=0.3]{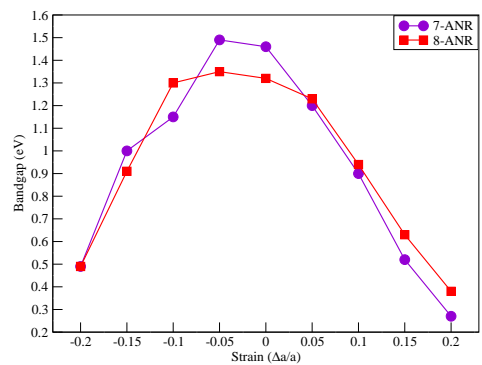}
 \caption{Energy gap of an A\MoSTwo{}NR-H-O under various strains for w= 7, 8\label{gapstrain78}}
\end{figure}

\begin{figure}[b!]
   \centering
   \subfigure[][]{
      \label{fig:7}
      \includegraphics[width=0.4\linewidth]{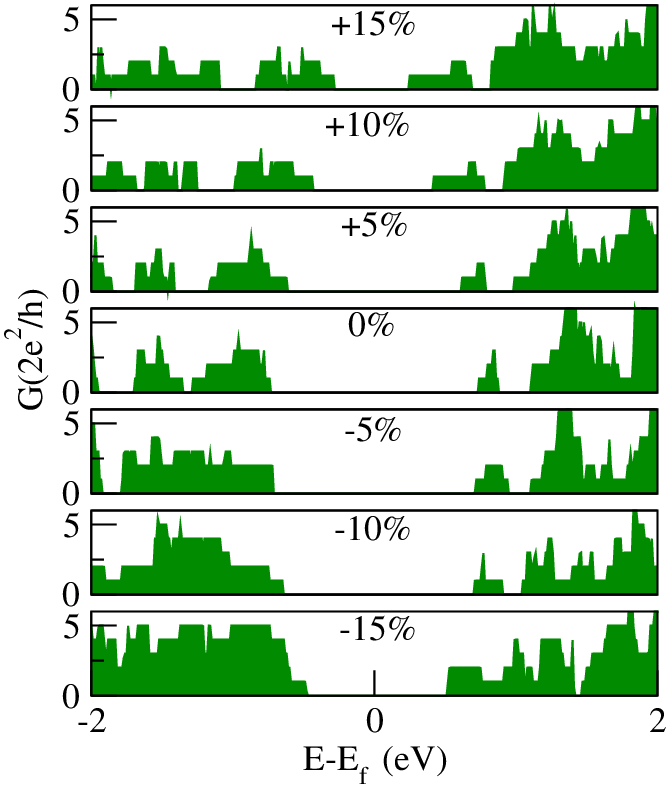}}
   \hspace{0.02\textwidth}
   \subfigure[][]{
      \label{fig:8}
      \includegraphics[width=0.4\linewidth]{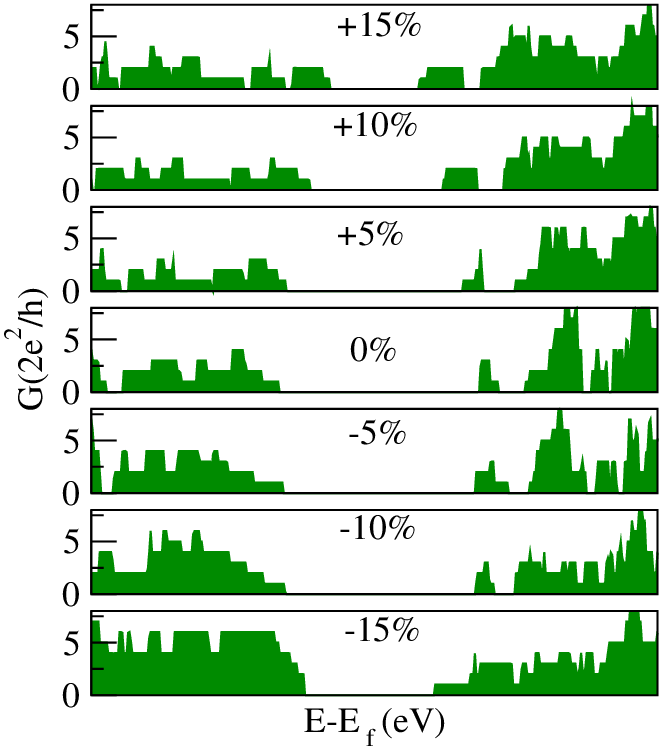}}
      \caption{The Landauer quantum conductance of an A\MoSTwo{}NR-H-O 
      in the y direction under various strains for w= 7, 8\label{transstrain78}}
\end{figure}

\vspace{2mm}
\textbf{Response of A\MoSTwo{}NR-H-Os to strain}: 

We investigated the effect 
of strain on A\MoSTwo{}NR-H-O. According to previous works\cite{Pan2012} we 
know that there are two typical families of A\MoSTwo{}NR-H-O, 
symmetric for odd width and asymmetric for even width. As a result,
A\MoSTwo{}NR-H-O with widths w = 7 and 8 was chosen to represent two 
typical families of A\MoSTwo{}NR-H-O. Now we want to extract band structure
in various strain and you can see our result in fig(\ref{gapstrain78}), with both 
positive and negative strain bandgap is decreased for both symmetric and 
asymmetric nanoribbons. With increasing or decreasing strain the number 
of passing channel for electrons is increased. 
Figure {\ref{transstrain78}} shows the quantum transport of 
the 7-,8-A\MoSTwo{}NR-H-O under various induced strain at T = 0 K. 


In room temperature, we can use Eq.~\ref{1} for computing quantum conductance 
in various strain for the 7-,8-A\MoSTwo{}NR-H-O as you see in Fig.~{\ref{G-W-S}}. 
With increasing or decreasing strain the number of passing channel for electrons 
is increased, but as you see in Fig.~{\ref{G-W-S}} in room temperature 
7-,8-A\MoSTwo{}NR-H-O have low conductance, as a result, you can use 
them for the transistor in high strain. Also if there is for example 1 V gate voltage
you can use these nanoribbons as strain sensors.

\section{Conclusion}
\label{sec:Conclusion}
Density functional studies of nanoribbons width and strain effects on the electrical 
transport properties of the  A\MoSTwo{}NR-H-O are presented. 
By applying a uniaxial tensile strain in the y-direction, the electronic properties 
of the A\MoSTwo{}NR-H-O nanoribbons were studied. 
Using the Wannier functions, the band structure and density of states were calculated for different strains from -15\% to +15\%. 
It is observed that A\MoSTwo{}NR-H-O in different width and different strain are a good candidate for using in the transistor. 
In addition, you can use them for strain sensors. 

\begin{acknowledgments}
We would like to acknowledge the Isfahan University of Technology. The authors gratefully 
acknowledge the Sheikh Bahaei National High Performance Computing Center (SBNHPCC) for 
providing computing facilities and time. SBNHPCC is supported by scientific and technological 
department of presidential office and Isfahan University of Technology (IUT). 
 
Discussions with \ldots about various choices of methods are appreciated.
\end{acknowledgments}
\appendix
\section{ANR-4-bare}
\label{app1}
\begin{figure}[b!]
 \includegraphics[width=\columnwidth]{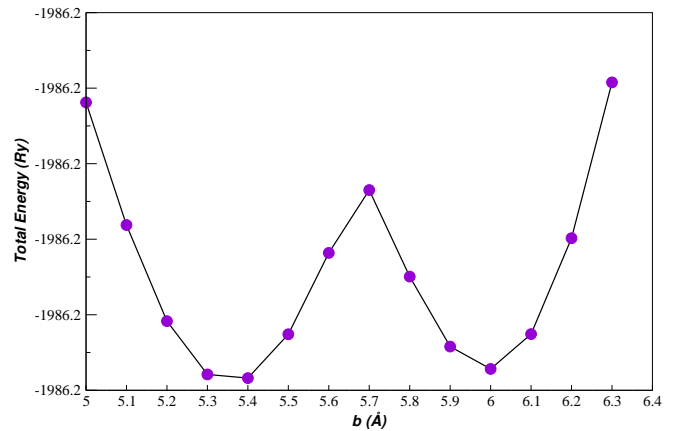}
 \caption{Total energy in terms of lattice constant for 4-A\MoSTwo{}NR-bare \label{2minimum}}
\end{figure}

\begin{figure}[ht]
 \includegraphics[width=0.35\textwidth]{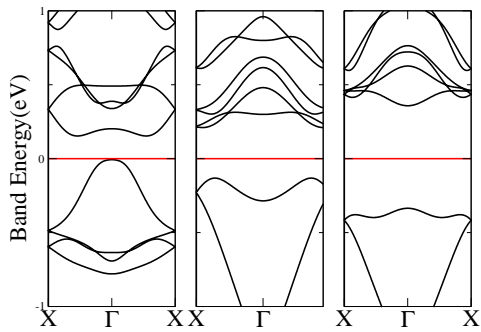}
 \caption{Band structure of 4-A\MoSTwo{}NR-bare for lattice parameter 5.4 , 5.7, and 6.0 respectively\angstrom{} \label{5.4}}
\end{figure}

We guessed that maybe there is a bond that reinforced in the second minimum in compare with 
maximum, so we checked that and we saw that (Fig.~\ref{2minimum}). 
But there was no bond that approves this guess. 
Then we investigate that if one of this minimum or maximum has magnetic 
structure, and the value of total magnetization is zero. 
We plot band structure for these 3 that you can see in Figs.(\ref{5.4}). 

We have seen that in the case of  4-A\MoSTwo{}NR-bare for lattice parameter 5.4 the bandgap is 0.15 eV (indirect) that is in good agreement with the previous result, 
and in the case of 4-A\MoSTwo{}NR-bare for lattice parameter 5.7,6.0 the band gap is 0.32eV (indirect) and 0.67eV (direct in $\Gamma$ point) respectively. 
\bibliography{my_bibliography}

\end{document}